Article

# Digital Distractions from the Point of View of Higher Education Students

María Ángeles Pérez-Juárez , David González-Ortega and Javier Manuel Aguiar-Pérez *

Departamento de Teoría de la Señal y Comunicaciones e Ingeniería Telemática, Universidad de Valladolid, ETSI Telecomunicación, Paseo de Belén 15, 47011 Valladolid, Spain; mperez@tel.uva.es (M.Á.P.-J.); davgon@tel.uva.es (D.G.-O.)
* Correspondence: javagu@tel.uva.es

**Abstract:** Technology enables a more sustainable and universally accessible educational model. However, technology has brought a paradox into students' lives: it helps them engage in learning activities, but it is also a source of distraction. During the academic year 2021–2022, the authors conducted a study focusing on classroom distractions. One of the objectives was to identify the main digital distractions from the point of view of students. The study was carried out at an engineering school, where technology is fully integrated in the classroom and in the academic routines of teachers and students. Discussions and surveys, complemented by a statistical study based on bivariate correlations, were used with participating students (n = 105). Students considered digital distractions to have a significant impact on their performance in lab sessions. This performance was mainly self-assessed as improvable. Contrary to other contemporary research, the results were not influenced by the year of study of the subject, as the issue is important regardless of the students' backgrounds. Professors should implement strategies to raise students' awareness of the significant negative effects of digital distractions on their performance, as well as to develop students' self-control skills. This is of vital importance for the use of technology to be sustainable in the long-term.

**Keywords:** higher education; educational technology; classroom distractions; digital distractions

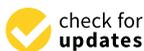

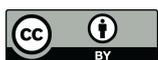

## 1. Introduction

Today's children and teenagers have grown up in classrooms where the use of different applications such as Classcraft, Code.org, Educaplay, Kahoot!, Genially, Pixton, RPGplayground, Storyjumper, among many others (e.g., [1–7]), was commonplace. Teachers and students of all educational levels are comfortable with technology, and devices and apps are commonly used to plan and develop learning activities. Our education system cannot do without technology [8]. Moreover, in general terms, the isolation during the COVID-19 pandemic and subsequent social distancing guidelines have increased the use of technology by instructors and students.

The presence of technology in classrooms opens the door to a huge range of opportunities that would not otherwise be possible. With the help of technology in the classroom, many educational benefits can be achieved. These benefits include inclusion (e.g., using assistive technologies for students with dyslexia, attention-deficit/hyperactivity disorder, or visual impairments), student engagement, facilitating instructor-student and student-student interactions, and creating interesting learning opportunities (e.g., [9–13]).

In addition, technology allows for a more sustainable and universally accessible educational model in many ways, including: (1) facilitate access to education, (2) reduce costs, (3) reduce carbon emissions and environmental impact, (4) enable personalized learning, and (5) increase collaboration. Technology can enable access to education for students who are unable to attend traditional education programs for many reasons including geographical or financial constraints. Online courses and virtual campuses can reach students in remote locations or those who cannot afford to attend a physical institution. Moreover, by using technology, educational institutions can reduce costs associated with physical infrastructure, transportation, and paper-based materials (printing). This promotes sustainable resource usage. Technology can also help reduce carbon emissions and other environmental impacts associated with traditional educational models. In doing so, technology can help educational centers become more environmentally sustainable. Technology can enable personalized learning, which allows students to learn at their own pace and according to their own preferences. This can reduce student dropout rates and make learning more sustainable over time. Technology can enable collaboration between students and professors from different parts of the world, fostering a more diverse and sustainable learning environment. By enabling international collaboration, technology can help students develop a global perspective and learn about sustainable practices in different cultures. Overall, technology can enable a more sustainable educational model by increasing access, reducing costs, and minimizing waste. By leveraging technology, educational institutions can promote sustainability and ensure that education is accessible to everyone, regardless of their location or financial situation.

Technological devices, including smartphones, tablets, and laptops, are ubiquitous on university campuses and in university classrooms. Most university students bring one or more of these devices into the classroom (e.g., [14–17]). College students apparently use these digital devices to take part more appropriately in learning activities. However, digital devices can also hinder the learning process by making distracted students think that they can multitask without affecting their academic performance. In fact, technology has brought a paradox into students' lives: technology helps them participate in learning activities, but technology is also a source of distraction from getting their tasks done. Nevertheless, eliminating technology from classrooms to prevent digital distractions would generate inequalities between students with and without access to technology outside the classroom. In addition, digital illiteracy could increase among the youngest learners [8]. For these reasons, it is important that both professors and students make appropriate use of technology so that its use in education is sustainable and can be maintained in the long-term.

Social media is fully integrated into college students' daily lives. Today's university students, frequently referred to as the Net Generation [18], commonly use technology [19]. In fact, according to scholars, university students spend a huge amount of time using their technological devices for off-task purposes. In [20], the authors conducted a survey which reported that 92% of university students used messaging apps on their smartphones during lectures. A few years later, another survey [21] found that for around 21% of class time university students used their digital devices for off-task purposes.

The authors are aware of the challenge of using technology in university classrooms, due to the risk of technology causing distractions for university students. For this reason, during the academic year 2021–2022, the authors conducted an educational innovation project focused on digital distractions. Among other objectives, the project addresses the identification of those that, from the point of view of students, are the main distractors associated with technology. This study on distractors was carried out at the Higher Technical School of Telecommunication Engineering at the authors' university in central Spain. This engineering school offers different engineering study programs related to Telecommunication Technologies to students aged approximately from around 17 to 25. A great number of students have been involved in this project. The participation of many students was considered a priority to adequately get the students' perception on digital distractions.

First, the need for sustainable use of technology in the classroom is briefly presented, as well as the challenges posed by this new scenario, focusing on instructors' technological skills and on digital distractions faced by students. Next, the existence of digital distractions in the classroom is analyzed with further detail. Then, the data collection process to identify classroom distractors is described, as well as other details of the study which are developed.

Afterwards, the results obtained are presented and discussed. Finally, the main conclusions are presented.

## 2. Background

*2.1. Necessity and Challenges of Sustainable Digital Uses in College Classrooms*

Over the last few years, university campuses have progressively been equipped with technology. In fact, the presence of technology on university campuses is expected to be increasingly noticeable, as many universities are undergoing digitization processes. Students use many different types of technological devices and software applications during academic activities.

There are different reasons and motivations for students to bring computers to university classes, including personal preference and the desire to use them for notetaking or information research [22].

Many professors are adapting their methodologies to incorporate the use of technology [23]. The use of technology on university campuses means that professors must face new and demanding challenges. Instructors must be technologically skilled, and able to efficiently integrate the use of technology into methodologies and into the learning activities planned. This is of vital importance for the use of technology to be sustainable in the long-term. In [24], the authors researched perceptions of Iranian English professors and other foreign languages college professors on using technology. These researchers confirmed that technological literacy was one of the main concerns of the instructors.

Therefore, it is important that professors accept the challenge of incorporating the use of technology in the classroom and dedicate time and effort to face this challenge successfully. Some scholars have researched how university professors acquire the knowledge and skills necessary to use technology in the classroom, finding that this process happens in many ways (e.g., [25]).

Monitoring students' use of technology during their work is also very important and necessary for a sustainable use of technology in the long-term. In [26], the authors note that laptops are widespread in university classrooms and that, although they are a valuable tool, they allow students to be distracted by the Internet. So, it is important to guarantee that students are not getting distracted by using technology for off-task purposes.

*2.2. Digital Distractions in College Classrooms*

Students frequently multitask during academic activities [27]. Although some think that multitasking is synonymous with efficiency, there is evidence that most students find multitasking to be very distracting (e.g., [28]). In fact, there are many researchers who have conducted studies dismantling the digital native myth (e.g., [19,27,29–37]). Contemporary research has demonstrated that the ability to multi-task does not exist. Activities take longer for students due to the time spent on distracting activities and the extra effort involved in getting back on task [38]. Attending lectures and engaging with digital technologies for off-task purposes at the same time has a negative impact on students' learning outcomes and academic performance. Students who regularly multitask for off-task purposes do not perform their academic tasks as well as those who only focus on their work. If the student's academic performance is affected, this situation will not be sustainable over time since it would mean a significant deterioration of the learning process and its results.

This is consistent with the cognitive bottleneck theory of attention, the selective attention theory, and the cognitive load theory.

The cognitive bottleneck theory of attention focuses on the limitations of attentional processing, suggesting that attention is a limited resource and that there is a bottleneck in information processing that limits the amount of information that can be processed at any given time. The theory suggests that this bottleneck occurs at the level of working memory and that the attentional selection process helps to prioritize and filter incoming information [39]. Digital distractions can easily overwhelm our attentional system, leading to reduced attentional resources for other tasks. Despite the capabilities offered by cognitive control, humans have a limited capacity to carry out several control-dependent processes simultaneously [8].

The selective attention theory focuses on how attention is selectively directed towards certain stimuli while ignoring others. This theory suggests that attentional selection is necessary to focus on relevant information and ignore irrelevant information. According to this theory, the brain has limited resources and must selectively allocate attention to important information to avoid overload [40]. The presence of technology and associated distractions can make it difficult for students to exercise selective attention and cause them to lose focus on important tasks.

The cognitive load theory focuses on the limitations of working memory and suggests that there is a limit to the amount of information that can be held and processed in working memory at any given time (e.g., [41–45]). The constant presence of technology and associated distractions can increase students' cognitive load, which could interfere with their ability to process and retain information.

In [46], the authors give a very visual and revealing description of a university classroom as a place where all heads are tilted towards a flickering screen. Digital overload is a common problem today. Students are constantly bombarded with messages and alerts on their laptops, tablets, and smartphones, which makes focusing on their tasks difficult for them. So when students are tempted to procrastinate, distractions and diversions are only one click away.

According to the social learning theory, a student's learning is influenced by their social environment [47]. Technology can create a culture of distraction in which students feel pressured to be constantly connected and distracted by incoming notifications and messages. This culture can cause students to lose their ability to focus and maintain their attention on a specific task. Additionally, the situated cognition theory asserts that knowledge is closely linked to the context in which it is acquired [48]. If students are constantly exposed to technology and distractions, they may be less likely to remember specific information because they are not connecting this knowledge with a meaningful context.

*2.3. Impact of Digital Distractions in College Classrooms*

Any stimulus or information that diverts an individual's attention from the main task at hand is considered a distractor. Distractors affect the human ability to concentrate and have an external (noise, movement, etc.) or internal (thoughts, feelings, etc.) origin [49].

When it is digital technology that diverts an individual's attention away from the primary task being done, it is referred to as a digital distractor (e.g., [50–52]).

Some instructors think that a student can only be distracted by another student. However, many of the interruptions that students experience are caused by their technological devices. In fact, an important number of researchers (e.g., [8,53]) remark that many classroom situations can be distractors, and that some of these distractions are caused by technology including email, instant messaging apps, social networks, and games.

There are several reasons why monitoring technology use and managing distractions is important for long-term sustainable and effective technology use, including: (1) improved concentration, (2) healthy habits, (3) better time management, (4) reduced cyberbullying and misuse, and (5) improved learning outcomes. Distractions, especially those resulting from technology, can hinder a student's ability to focus on their studies. By supervising technology use and minimizing distractions, students are more likely to focus on their studies, leading to improved academic performance. In addition, excessive use of technology can lead to addiction and harmful habits, which can negatively affect physical and mental health. Professors can help students to learn healthy habits for using technology in a sustainable way. Unsupervised technology use can lead to wasted time and reduced productivity. By supervising technology use, professors can help students manage their time effectively, and teach them to use technology in a way that supports their learning goals. Unsupervised technology use can also lead to cyberbullying and other forms of harmful behavior. Supervising the use of technology can reduce the risk of this type of behavior and promote a safe and respectful online environment. In addition, by monitoring technology use, professors can ensure that students are using technology in ways that enhance their learning experience. This includes the use of technology for research, collaboration, and other educational purposes. Overall, monitoring technology use and managing distractions are crucial to sustainable and effective use of technology in education. By promoting healthy habits and minimizing distractions, professors can help students use technology in a way that supports their academic success and overall well-being.

In [54], the authors have identified three different lines of research regarding digital distractions and their consequences. Firstly, some researchers have focused on providing insight into what students do when they are using their technological devices for off-task purposes during academic activity (e.g., [20,26,55–59]). Secondly, some scholars have researched the negative consequences of digital distractions on students' learning outcomes and academic performance (e.g., [46,60–68]). Finally, some researchers have focused on identifying digital distractions' determinants, for example, gender or age, and the extent to which these determinants influence digital distractions (e.g., [69–72]). Some of these studies and surveys are briefly described below.

Regarding the research on the use of technology for off-task purposes, the authors in [55] tried to identify the types of applications commonly used by students in class, and their reasons for doing so. Results showed that students regularly used technology for a variety of non-academic reasons, and that text messaging and email were the most commonly used applications.

The study conducted by the authors in [56] found that during class, university students often engaged in using online communication, online photo sharing, and online information seeking, among other usages.

In [57], the author asked college students about their perceptions regarding the use of digital devices for off-task purposes during lessons. Students reported an average use of technology for off-task purposes of 10.93 times per day, and 80% of them admitted that this behavior distracted from their duties and tasks.

In [58], the authors used in-class observations and surveys to analyze the use of laptops by students. These authors observed changes in computer use during an unmonitored 165-min class without restrictions. Student reports provided similar information on laptop activities as were provided by the observations. Notetaking and social media websites were the most common uses of laptops. The data showed that students spent almost two thirds of their time on off-task computer activities. It is also remarkable that students' engagement in their tasks had great variations throughout the class.

Regarding the negative impact of digital distractions on students' learning outcomes and academic performance, the experimental study conducted in [60] demonstrated lower performance in a lecture quiz of those students who did off-task text messaging during the lecture versus those who did not.

Moreover, the research presented in [26] showed that the class performance of students who brought laptops and frequently used Internet for off-task purposes was affected negatively.

On their side, the experiment presented in [62] showed that divided attention between technology and a lecture did not have a negative impact on the comprehension of this lecture but significantly reduced long-term retention of the lecture. Results were measured through a within-class quiz and unit and final exams performance, respectively.

The experiment described in [64] on the use of mobile phones in a classroom revealed that message creation unrelated to a class lecture negatively impacted students' learning outcomes.

In [66], the author explored students' disordered use of technology. His study remarked on different types of digital inconvenience, ranging from short interruptions to the ways in which technology was significantly diminishing students' work.

In terms of determinants influencing digital distractions, the research presented in [70] showed that impulsive individuals are more prone to digital distractions, and that habitual use of technology is the strongest determinant of digital distraction intensity. In fact, accord-ing to these authors, the intensity of digital distraction is determined by an individual's Internet addiction; i.e., Internet-addicted students are more prone to digital distractions during academic tasks.

In [72], the authors examined how media multitasking affects students' social and psychological well-being. They found that media multitasking involved different behaviors with differing effects (null, negative, and positive) on students' perceptions of psychological and social wellbeing.

It is also important to remark that students' classroom technology use for off-task purposes can distract their peers and negatively affect the learning outcomes of neighboring students. This was demonstrated in the experiment conducted by the authors in [17] where students scored better on post-conference questions about contents covered while neighboring students were on-task note taking than on those questions about contents covered while neighboring students were off-task web browsing, irrespective of where they sat in relation to the classmate.

Keeping students motivated, thinking, doing activities, and far away from boredom are the best ways to prevent digital distractions [73]. In fact, research has revealed that students' attention increases during and immediately after a change in behavior or pedagogy of the instructor (e.g., [74,75]). However, even considering that student involvement and active learning are key to overcoming distractions, distractions remain present in a significant way.

Finally, it is also interesting to notice that a bit of cyberloafing as leisure can be positive when performed moderately as a means of recovery outside the classroom. In the study presented in [68] involving 1050 undergraduate students at a large Chinese University, the test results showed a negative relation between academic performance and media multitasking in class. However, the relationship between academic performance and cyberloafing out of class was inverted U-shaped.

## 3. Methods

### 3.1. Settings

This study on classroom distractors has been carried out at the Higher Technical School of Telecommunication Engineering of the University of Valladolid, an engineering school that offers different engineering study programs related to Telecommunication Technologies, to students aged approximately from around 17 to 25, and where technology is fully integrated into the learning activities and methodologies used by instructors in different courses. It is also important to mention that both instructors and students are regular users of technology, feel comfortable using technology, and consider it a necessary part of their lives. This total integration of technology in the academic lives of students and professors is probably a consequence of the types of studies offered at this educational center.

### 3.2. Participants

Table 1 shows the subjects participating in this study. The subjects reflected in this study address different areas including programming languages and paradigms, image and sound fundamentals, technologies, programming languages, and tools for the development of different types of software systems such as distributed applications or web applications, transmission networks, and telematics services. This study included the following participating subjects: Programming (PRO), Wired and Wireless Transmission Networks (RTCI), Distributed Application Development (DAD), Technologies for Web Applications (TAW), Image and Sound Fundamentals (FIS), Telematic Systems Development Laboratory (LDST), and Integration of Telematics Services in Next Generation Networks (ISTR).

**Table 1.** Subjects participating in the study about classroom distractors, including year in which the subject is taught, the engineering study program to which the subject belongs, the number of students enrolled in the subject, and the number of respondent students.

| Subject | Year | Engineering Studies | Enrolled Students | Respondent Students |
|---|---|---|---|---|
| PRO | 1st | GITT-GITET | 31 | 31 |
| RTCI | 3rd | GITET-T | 16 | 16 |
| DAD | 3rd | GITT | 34 | 26 |
| TAW | 4th | GITET-T | 11 | 9 |
| FIS | 4th | GITT | 23 | 11 |
| LDST | 4th | GITT | 8 | 5 |
| ISTR | 5th | Master | 8 | 7 |
| Total | | | 131 | 105 |

For each subject, Table 1 details the following data: (1) year in which the subject is taught, (2) engineering study program to which the subject belongs, (3) number of students enrolled in the subject, and (4) number of respondent students (students present in the laboratory on the day the study was carried out).

Furthermore, as shown in Table 1, except for ISTR, the subjects are part of one, or both, of the following Telecommunication Engineering study programs: (1) Bachelor's Degree in Telecommunication Technology Engineering (GITT), and (2) Bachelor's Degree in Specific Telecommunication Technology Engineering (GITET). The latter has three specialties which are Telematics (GITET-TEL), Telecommunication Systems (GITET-ST), and Electronic Systems (GITET-SE). ISTR is part of the Master's Degree in Telecommunication Engineering that can be studied after having completed any of the Bachelor's Degrees.

When the number of students enrolled in a subject and the number of respondent students do not match, this is because some students were absent or did not take part in the study for different reasons, including illness, carrying out their tasks elsewhere, or having abandoned the task and/or the subject.

As shown in Table 1, the number of participating students was 105. The participating subjects were chosen in order to reach a large part of the students enrolled in the Higher Technical School of Telecommunication Engineering while avoiding repetitions as far as possible. In other words, the goal was that each student would participate in the discussion on digital distractions only once.

### 3.3. Identification of Classroom Distractors

In [53], the authors attempted to identify the main distractions in the classroom from the students' point of view. These researchers identified 17 situations produced in the classroom by the students themselves and 24 situations produced externally. The number of distractors identified by these authors was very comprehensive and included both digital and non-digital distractors of different natures.

The list of distractors proposed by these authors was completed and adapted to the context of the Higher Technical School of Telecommunication Engineering of the University of Valladolid where our study to identify classroom distractors was conducted.

The complete list of possible distractors is shown in Table 2 in Section 4.1 Results: Classroom Distractors. The items were classified into two categories: non-digital and digital distractors. Digital distractors are at the end and have been marked with an asterisk (*).

### 3.4. Procedure

The objective of the study conducted was to identify the main distractors in the context of laboratory classes taken by students, which is where students have greater freedom to plan their own activity and use of time. During the development of these laboratory classes, professors usually present the activity to be carried out at the beginning of the class and, from that moment on, it is the students themselves who decide how to manage their time and the use of the technological resources available.

The survey and subsequent discussion about classroom distractions took place during a mid-semester laboratory class. During this laboratory class, the students carried out the scheduled activity according to the course work plan, but about 30 min were reserved for the survey and discussion on classroom distractions.

The professor explained aloud to the students present the purpose of the study. The students were also informed that the study was being developed within the framework of a teaching innovation project supported by the Vice-Rectorate for Teaching Innovation and Digital Transformation of the University, and why it is important for the University to promote the participation of professors and students in teaching innovation projects.

Students were then provided with the list of digital and non-digital distractors shown in Table 2 in Section 4.1 Results: Classroom Distractors and asked to point out those that, in their opinion, had the greatest impact. Students were also asked to identify classroom distractions other than those proposed by their professors.

In addition, the students were asked: (1) if they thought they could make better use of their time during the laboratory classes; and (2) if, in general terms, they had the perception that digital distractors had more weight than non-digital ones.

For data collection, most professors opted for a survey through the subject website offered on the University's Moodle-based Virtual Campus. However, there was also the case of some professors who preferred to prepare the survey on paper. This decision regarding data collection was made by the professor responsible for the subject. Each professor chose to collect the data in the way he or she considered most appropriate. Regardless of whether it was conducted through the University Moodle-based Virtual Campus or on paper, the survey was completely anonymous. Students were informed of this fact before taking the survey and were encouraged to provide their answers freely and honestly.

Data collection on classroom distractions was followed by a brief discussion (focus group-style conversation) in which students were encouraged by a professor to make any comments they considered appropriate on the topic in question in order to collect those that might be of interest for the planning of the subjects. Each discussion on classroom distractions was moderated by a professor and carried out in small groups, which allowed the professors to comment on different aspects with the students. The number of participating students in the discussion on classroom distractions in each subject is detailed in Table 1.

## 4. Results

### 4.1. Classroom Distractors

Table 2 shows the list of classroom distractors proposed to the students, and the number and percentage of students who considered each of the proposed distractors relevant. Global numbers and percentages are presented, as well as numbers and percentages broken down by subject.

**Table 2.** Influence of classroom distractors (digital and non-digital) from the students' point of view. Presents global figures and percentages, and figures and percentages broken down by subject. Digital distractors are at the end and have been marked with an asterisk (*).

|  | PRO | PRO% | RCTI | RTCI% | DAD | DAD% | TAW | TAW% | FIS | FIS% | LDST | LDST% | ISTR | ISTR% | Total | Total% |
|---|---|---|---|---|---|---|---|---|---|---|---|---|---|---|---|---|
| Students talking to others or alone. | 7 | 22.58% | 9 | 56.25% | 14 | 53.85% | 8 | 88.89% | 4 | 36.36% | 3 | 60.00% | 4 | 57.14% | 49 | 50.00% |
| Students asking non-important questions. | 6 | 19.35% | 1 | 6.25% | 4 | 15.38% | 2 | 22.22% | 2 | 18.18% | 0 | 0.00% | 1 | 14.29% | 16 | 16.33% |
| Students making unnecessary comments. | 9 | 29.03% | 10 | 62.50% | 8 | 30.77% | 3 | 33.33% | 4 | 36.36% | 1 | 20.00% | 1 | 14.29% | 36 | 36.73% |

**Table 2.** *Cont.*

| | PRO | PRO% | RCTI | RTCI% | DAD | DAD% | TAW | TAW% | FIS | FIS% | LDST | LDST% | ISTR | ISTR% | Total | Total% |
|---|---|---|---|---|---|---|---|---|---|---|---|---|---|---|---|---|
| Students working on homework for other courses. | 1 | 3.23% | 0 | 0.00% | 3 | 11.54% | 3 | 33.33% | 1 | 9.09% | 0 | 0.00% | 1 | 14.29% | 9 | 9.18% |
| Students wearing unusual clothes. | 0 | 0.00% | 1 | 6.25% | 2 | 7.69% | 0 | 0.00% | 0 | 0.00% | 0 | 0.00% | 0 | 0.00% | 3 | 3.06% |
| Students who make repetitive movements (touching their hair, cracking their knuckles, tapping their fingers, tapping their pencil, etc.). | 4 | 12.90% | 4 | 25.00% | 3 | 11.54% | 3 | 33.33% | 1 | 9.09% | 0 | 0.00% | 2 | 28.57% | 17 | 17.35% |
| Symptoms of illness of a student. | 4 | 12.90% | 1 | 6.25% | 4 | 15.38% | 3 | 33.33% | 4 | 36.36% | 0 | 0.00% | 2 | 28.57% | 18 | 18.37% |
| Students leaving early or arriving late. | 9 | 29.03% | 4 | 25.00% | 4 | 15.38% | 3 | 33.33% | 1 | 9.09% | 1 | 20.00% | 1 | 14.29% | 23 | 23.47% |
| Students leaving or entering. | 1 | 3.23% | 4 | 25.00% | 0 | 0.00% | 1 | 11.11% | 3 | 27.27% | 0 | 0.00% | 1 | 14.29% | 10 | 10.20% |
| Students eating or drinking. | 0 | 0.00% | 1 | 6.25% | 0 | 0.00% | 0 | 0.00% | 2 | 18.18% | 0 | 0.00% | 1 | 14.29% | 4 | 4.08% |
| Students yawning, nodding, or falling asleep. | 3 | 9.68% | 1 | 6.25% | 4 | 15.38% | 2 | 22.22% | 1 | 9.09% | 0 | 0.00% | 0 | 0.00% | 11 | 11.22% |
| Professor difficult to understand. | 8 | 25.81% | 11 | 68.75% | 11 | 42.31% | 4 | 44.44% | 7 | 63.64% | 5 | 100.00% | 2 | 28.57% | 48 | 48.98% |
| Professor doing unusual or repetitive movements. | 2 | 6.45% | 0 | 0.00% | 7 | 26.92% | 1 | 11.11% | 0 | 0.00% | 0 | 0.00% | 0 | 0.00% | 10 | 10.20% |
| Professor utilizing repetitive phrases or words. | 4 | 12.90% | 4 | 25.00% | 9 | 34.62% | 3 | 33.33% | 2 | 18.18% | 1 | 20.00% | 2 | 28.57% | 25 | 25.51% |
| Professor wearing unusual clothes. | 0 | 0.00% | 0 | 0.00% | 0 | 0.00% | 0 | 0.00% | 0 | 0.00% | 0 | 0.00% | 0 | 0.00% | 0 | 0.00% |
| Furniture (for example, broken or dirty chairs, tables, etc.,). | 6 | 19.35% | 3 | 18.75% | 2 | 7.69% | 3 | 33.33% | 2 | 18.18% | 1 | 20.00% | 0 | 0.00% | 17 | 17.35% |
| Problems with equipment (computers, keyboards, mice, etc., that work badly). | 8 | 25.81% | 10 | 62.50% | 13 | 50.00% | 7 | 77.78% | 6 | 54.55% | 4 | 80.00% | 3 | 42.86% | 51 | 52.04% |
| Climate (too hot/cold). | 14 | 45.16% | 8 | 50.00% | 12 | 46.15% | 6 | 66.67% | 11 | 100.00% | 5 | 100.00% | 6 | 85.71% | 62 | 63.27% |
| Students playing games on any device (smartphone, tablet, laptop, etc.).* | 6 | 19.35% | 12 | 75.00% | 8 | 30.77% | 6 | 66.67% | 1 | 9.09% | 4 | 80.00% | 3 | 42.86% | 40 | 40.82% |
| Students listening to music on any device (mp3 player, smartphone, tablet, laptop, etc.). * | 2 | 6.45% | 4 | 25.00% | 5 | 19.23% | 1 | 11.11% | 1 | 9.09% | 0 | 0.00% | 0 | 0.00% | 13 | 13.27% |
| Students using social networks on any device (smartphone, tablet, laptop, etc.). * | 6 | 19.35% | 13 | 81.25% | 12 | 46.15% | 6 | 66.67% | 2 | 18.18% | 0 | 0.00% | 4 | 57.14% | 43 | 43.88% |
| Students using messaging applications on any device (smartphone, tablet, laptop, etc.) for off-task purposes. * | 6 | 19.35% | 13 | 81.25% | 10 | 38.46% | 5 | 55.56% | 2 | 18.18% | 0 | 0.00% | 5 | 71.43% | 41 | 41.84% |
| Students using email (read or send) on any device (smartphone, tablet, laptop, etc.) for off-task purposes. * | 0 | 0.00% | 5 | 31.25% | 5 | 19.23% | 2 | 22.22% | 0 | 0.00% | 0 | 0.00% | 1 | 14.29% | 13 | 13.27% |

**Table 2.** *Cont.*

| | PRO | PRO% | RCTI | RTCI% | DAD | DAD% | TAW | TAW% | FIS | FIS% | LDST | LDST% | ISTR | ISTR% | Total | Total% |
|---|---|---|---|---|---|---|---|---|---|---|---|---|---|---|---|---|
| Students web browsing on any device (smartphone, tablet, laptop, etc.) for off-task purposes. * | 7 | 22.58% | 12 | 75.00% | 4 | 15.38% | 6 | 66.67% | 2 | 18.18% | 1 | 20.00% | 4 | 57.14% | 36 | 36.73% |
| Sounds produced by any device (smartphone, tablet, laptop, etc.) for different reasons (calls, notifications).* | 11 | 35.48% | 13 | 81.25% | 10 | 38.46% | 6 | 66.67% | 7 | 63.64% | 5 | 100.00% | 3 | 42.86% | 55 | 56.12% |

*4.2. Additional Distractors*

Students were also asked to identify other classroom distractors different from the ones proposed by the professors. Most of the students thought that the list of classroom distractors was very complete and contained most of the relevant distractors. However, students of the RTCI subject in the 3rd year identified an additional distractor that was insects coming inside the laboratory.

The classroom distractor related to insects was also pointed out by students of the PRO subject in the 1st year. The students of this course also pointed out two additional classroom distractors related to noise. These additional distractors were (1) noises from outside due to having the windows open, and (2) noises from the hallway due to having the door open.

*4.3. Use of Time*

In addition to asking the students to select the main distractors during the development of laboratory classes, the students were asked if they thought they could make better use of their time during the laboratory classes.

Table 3 shows the percentage of students in each subject that answered affirmatively to the question "How many students think they could better use their time during laboratory classes?", and Figure 1a shows its corresponding box plot.

**Table 3.** Percentage of students in each subject who thought that they could make better use of their time during laboratory classes. The year and the engineering study program to which the subject belongs are included.

| Subject | Year | Engineering Studies | Positive Answer |
|---|---|---|---|
| PRO | 1st | GITT-GITET | 77.42% |
| RTCI | 3rd | GITET-T | 100.00% |
| DAD | 3rd | GITT | 26.92% |
| TAW | 4th | GITET-T | 88.89% |
| FIS | 4th | GITT | 45.45% |
| LDST | 4th | GITT | 60.00% |
| ISTR | 5th | Master | 85.71% |
| Total | | | 69.20% |

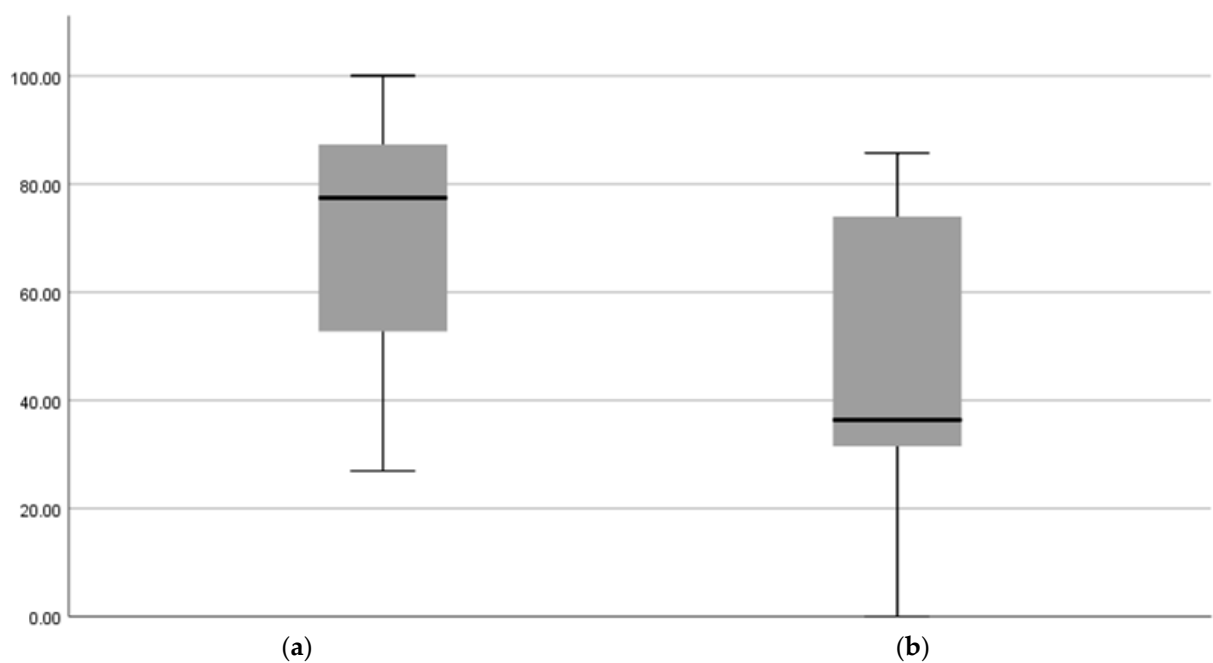

**Figure 1.** (**a**) Box plot of the results of the question "How many students think they could better use their time during laboratory classes" in the different subjects; (**b**) Box plot of the results of the question "How many students think that digital distractions have more influence than non-digital distractions" in the different subjects.

*4.4. Digital vs. Non-Digital Distractors*

In addition to asking the students to select the main distractors during the development of laboratory classes, the students were asked if, in general terms, they had the perception that digital distractors had more weight than non-digital ones.

Table 4 shows the percentage of students in each subject who thought that the digital distractions had more influence than the non-digital distractions, and Figure 1b shows its corresponding box plot.

**Table 4.** Percentage of students in each subject who thought that the digital distractions had more influence than the non-digital distractions. The year and the engineering study program to which the subject belongs are included.

| Subject | Year | Engineering Studies | Positive Answer |
| --- | --- | --- | --- |
| PRO | 1st | GITT-GITET | 32.26% |
| RTCI | 3rd | GITET-T | 81.00% |
| DAD | 3rd | GITT | 30.77% |
| TAW | 4th | GITET-T | 66.67% |
| FIS | 4th | GITT | 36.36% |
| LDST | 4th | GITT | 0.00% |
| ISTR | 5th | Master | 85.71% |
| Total | | | 47.46% |

## 5. Discussion

*5.1. Classroom Distractors*

The following are the main comments regarding the results presented in Table 2. Among the non-digital distractors, the students highlighted (1) the distractions caused by other students, (2) the distractions caused by the professor, and (3) the distractions related to the laboratory equipment and climate conditions.

Regarding the distractions caused by students, it is worth highlighting the distractions caused by students talking (alone or with other students, for example, with the student seated next to them) (50.73% of the students), or the distractions caused by students making unnecessary comments (36.73% of the students). These distractions are similar, as in both cases, students are having off-task conversations and/or doing off-task comments.

Regarding the distractions caused by the professors, it seems that not understanding professors' explanations has a special impact on students' concentration (48.98% of the students). It goes without saying that professors must make their explanations as easy to understand as possible and supervise their students' attention and performance. The professors must adapt to their students to optimize their learning and achievement, thus reducing the impact of all types of distractions.

It is also worth noting the distractions caused by the equipment used in the laboratory (screens, keyboards, mice, etc.) when they do not work properly (52.04% of the students), and the distractions caused by unsuitable climate conditions (too hot or too cold) (63.27% of the students), reaching even 100% of the students in the case of the 4th year FIS and LDST subjects. This is something important to consider since these situations can be easily corrected by the educational center. In this sense, it is important to establish adequate communication channels, so that equipment breakdowns, and those related to climate conditions, are solved as soon as possible, and do not interfere with student concentration and performance. Concerning the distractions caused by unsuitable climate conditions, the special conditions caused by the COVID-19 pandemic must be mentioned given that the University of Valladolid established different regulations, such as to keep the doors and windows open as long as possible to minimize exposure to fine aerosol particles, thus preventing COVID-19 infections from happening.

On the other hand, among the digital distractions, it is worth highlighting social networks (43.88% of the students) and messaging applications (41.84% of the students) as the main distractors with quite similar percentages, well ahead of email (13.27% of the students) or music (13.27% of the students).

Regarding music, it should be noted that in the laboratory classes of certain subjects, some students ask the professor for permission to listen to music. This happens for example in laboratories related to coding and when using different languages and programming paradigms, especially when students are working individually and not in a team. According to the students themselves, the reason for doing this is that music at an adequate low volume can favor concentration, as it helps students to isolate themselves from a noisy and/or distracting environment.

On the other hand, regarding email, the result was expected, since in recent times email has lost weight compared to social networks and messaging applications used to socially interact with others. This is the reason why students pay less attention to email comparing to social networks and messaging applications. In general, students consider that information, images, documents, etc., received through email are probably not important enough to capture their immediate attention. Even in the case of messages sent by their university or their professors, messages are not generally urgent, and students feel they can delay their reading until the end of the laboratory class.

Another relevant distractor is the sounds produced by different devices (smartphones, tablets, laptops, etc.) for different reasons (calls, notifications, typing on keyboards, etc.) (56.12% of the students). In this case, the result obtained was a surprise for the authors who did not expect such a high impact. Fortunately, in some cases, this is something that can be easily solved by asking students to mute the sounds of their devices during laboratory classes. In fact, all the professors participating in the project plan to consider this for the future, since it seems that even at a low volume, the sounds emitted by technological devices are annoying, distract the students, and have an impact on their concentration level.

Some researchers have focused on the main determinants for digital distractions. For example, according to [69], males check their technological devices in classroom more often than females. Furthermore, according to [70], students' age is negatively associated with digital distraction; older students are less prone to digital distraction.

Regarding the present study, the authors decided not to take gender into account, for several reasons. First, the percentage of women in the engineering study programs where the study was conducted is very low, with some subjects only attended by men, or with the presence of a single woman. Second, prior to the beginning, the perception of the professors participating in the study was that gender is not a big determinant of the behavior of students, not only speaking in general terms, but also in regards to classroom distractions, and that there are other factors that could have more impact such as age, or some personality traits, e.g., impulsivity. Third, the researchers responsible for this study felt that students could be disappointed if a distinction was made by gender in a scenario (a learning process) where gender is the least important.

On the other side, the age of the student was not among the data collected within the study, however, the year of the engineering study program in which each course is taught was registered (see Table 1). In this case, the results obtained do not support the conclusions pointed out by [68], as in our case, it is not obvious that the later a subject is taken, the fewer the distractions. In fact, the percentage of students influenced in the 1st year subject PRO is not significantly higher, or is even lower, than the percentage of students influenced in later year subjects, for different digital and non-digital distractors.

*5.2. Additional Distractors*

Students were also asked to identify other classroom distractors different from the ones proposed by the professors. Some students (RTCI subject in the 3rd year and PRO subject in the 1st year) identified an additional distractor that was insects coming inside the laboratory. This happens when windows are open, which was usual due to the COVID-19 pandemic, as bringing fresh outdoor air into the classrooms helped to avoid virus particles from accumulating inside, and consequently reduced professors and students SARS-CoV-2 exposure risk.

PRO subject in the 1st year also pointed out two additional classroom distractors related to noise, and in some ways also a consequence of the COVID-19 pandemic. These additional distractors were (1) noises from outside due to having the windows open, and (2) noises from the hallway due to having the door open. As explained previously, windows and doors were usually kept opened, as recommended by the University, to reduce professors' and students' SARS-CoV-2 exposure risk. This situation improved the quality of air in the laboratories, but unfortunately created other problems related to noise coming from outside the laboratory, and to climate conditions especially in very cold/hot months. It is also worth mentioning once again that noise was revealed to be quite distracting for students, including all types of noise, such as noises generated by technological devices (smartphones, tablets, laptops, etc.) due to different reasons (calls, notifications, typing in keyboards, etc.) coming from inside the laboratory, but also noises coming from outside such as human voices or traffic noise.

*5.3. Use of Time*

Students were also asked if they thought they could make better use of their time during the laboratory classes.

According to the results shown in Table 3 and Figure 1a, a large percentage of students on average (69.20%) thought they could make better use of their time during the development of the laboratories. The obtained standard deviation is 26.18. There is no trend in the results as a function of the year of the subject. In this regard, the percentage of students who answered affirmatively in both the subject of the first year (PRO) and the subject of the Master (ISTR) is quite large and above the average, 77.42% and 85.71%, respectively. It seems that the perception of the students about their improvable performance in the laboratory classes persists throughout the years, and it is mainly influenced by the characteristics of each subject. In the case of some subjects, such as the 3rd year subject RTCI, the percentage reached 100%. In other subjects, however, the percentage was lower, for example, 26.92% in DAD. However, the reason for this low percentage is that, in the case of this subject, attendance for the laboratory class was not compulsory. The students themselves explained to the professor that they preferred to be as diligent as possible, to finish the tasks during the laboratory classes and avoid homework.

*5.4. Digital vs. Non-Digital Distractors*

Students were also asked if, in general terms, they had the perception that digital distractors had more weight than non-digital ones.

According to the results shown in Table 4 and Figure 1b, a significant percentage of students on average (47.57%) thought that digital distractions have more influence than non-digital ones, with a standard deviation of 31.24. It should be noted that in the list of classroom distractors provided to the students, non-digital distractors were considerably more numerous than digital distractors. However, almost half of the students felt that distractions caused by technology outweighed those caused by other reasons. Similar to the results of the question about the students' performance in the laboratory classes, there is no trend in the results as a function of the year of each subject.

It is remarkable that in the subject belonging to the Master (ISTR), the students thought that the digital distractions were more influential than the non-digital distractions (85.71% of the students), while in the subject of the first year (PRO), the number of students that gave more weight to the digital distractions was relatively small (32.26% of the students). This result could have been influenced by the immaturity of first year students, which could lead them to think that digital distractions do not affect them as much as they really do, because their continuous use of technological devices prevents them from having a critical awareness of the inappropriate use of these devices. In the authors' opinion, it is likely that this tendency regarding the increasing importance of digital distractions will continue, and even increase in the future, due to the digitalization process of college classrooms in university campuses.

*5.5. In-Depth Discussion of Results*

For an in-depth analysis of the results, the authors have accomplished a statistical study consisting in computing bivariate correlations between the different distractions and the questions "How many students think they could better use their time during laboratory classes?" and "How many students think that digital distractions have more influence than non-digital distractions?". For each distraction, the variable considered in the study has as samples the percentage of students in each subject who considered the distraction as relevant. The distraction "Professor wearing unusual clothes" was excluded from the analysis as no student in any subject selected it. First, the Shapiro-Wilk test [76] was used to check if the variables follow a normal distribution. All the variables, except "Students wearing unusual clothes" and "Professor doing unusual or repetitive movements", follow a normal distribution. The Pearson correlation coefficient can be used if at least one of the two variables to be correlated follows a normal distribution [77]. Therefore, the Pearson correlation coefficient was calculated for all pairs of variables, except for the correlation in which both the variables "Students wearing unusual clothes" and "Professor doing unusual or repetitive movements" took part. In these cases, the Spearman correlation coefficient was used.

After calculating the correlation coefficients between each pair of variables, their statistical significance was tested by computing the $p$-value. When the $p$-value is lower than 0.05, the null hypothesis is rejected, and the statistical significance of the correlation value is verified.

Table 5 shows the correlation coefficient matrix of the digital and non-digital distractions and the two questions. The correlation coefficients with statistical significance are highlighted in bold, distinguishing between the *p*-values lower than 0.05 and lower than 0.01. The question "How many students think they could better use their time during laboratory classes?" is only correlated significantly with the distraction "Students web browsing on any device (smartphone, tablet, laptop, etc.) for off-task purposes" with a Pearson coefficient of 0.878. The larger the number of students who have selected this distraction in a subject, the larger the percentage of students who think they could better use their time during laboratory classes. Therefore, this digital distraction stands out as having the most influence on the performance of students in laboratory classes, and it would be important to establish measures to limit its impact together with reinforcing the self-control of students to improve the learning process in laboratory classes.

**Table 5.** Correlation coefficient matrix of the different distractions and the questions "How many students think they could better use their time during laboratory classes?" and "How many students think that digital distractions have more influence than non-digital distractions?".

|  | How Many Students Think They Could Better Use Their Time During Laboratory Classes? | How Many Students Think That Digital Distractions Have More Influence Than Non-Digital Distractions? |
|---|---|---|
| Students talking to others | 0.283 | 0.324 |
| Students asking non-important questions | −0.046 | 0.285 |
| Students making unnecessary comments | 0.293 | 0.335 |
| Students working on homework for other courses | 0.118 | 0.383 |
| Students wearing unusual clothes | −0.247 | 0.126 |
| Students who make repetitive movements | 0.691 | 0.923 ** |
| Symptoms of illness of a student | −0.040 | 0.409 |
| Students leaving early or arriving late | 0.624 | 0.164 |
| Students leaving or entering | 0.339 | 0.585 |
| Students eating or drinking | 0.003 | 0.380 |
| Students yawning, nodding, or falling asleep | −0.124 | 0.084 |
| Professor difficult to understand | −0.132 | −0.475 |
| Professor doing unusual or repetitive movements | −0.548 | −0.180 |
| Professor utilizing repetitive phrases or words | −0.059 | 0.418 |
| Furniture | 0.274 | −0.159 |
| Problems with equipment | 0.077 | −0.119 |
| Climate (too hot/cold) | −0.157 | −0.243 |
| Students playing games on any device | 0.507 | 0.135 |
| Students listening to music on any device | 0.009 | 0.302 |
| Students using social networks on any device | 0.546 | 0.887 ** |
| Students using messaging applications on any device for off-task purposes | 0.609 | 0.953 ** |
| Students using email on any device for off-task purposes | 0.413 | 0.715 |
| Students web browsing for off-task purposes | 0.878 * | 0.870 * |
| Sounds produced by any device | 0.201 | −0.226 |

* Correlation is significant at *p* < 0.05 level; ** Correlation is significant at *p* < 0.01 level.

Meanwhile, there are four distractions correlated significantly with the question "How many students think that digital distractions have more influence than non-digital distractions?", namely the non-digital "Students who make repetitive movements (touching their hair, cracking their knuckles, tapping their fingers, tapping their pencil, etc.)" (Pearson coefficient of 0.923), and the three digital "Students using social networks on any device (smartphone, tablet, laptop, etc.)" (Pearson coefficient of 0.887), "Students using messaging applications on any device (smartphone, tablet, laptop, etc.) for off-task purposes" (Pearson coefficient of 0.953), and "Students web browsing on any device (smartphone, tablet, laptop, etc.) for off-task purposes" (Pearson coefficient of 0.870). The last three digital distractions, which have been selected as relevant by a large percentage of students, stand out as the most influential for the relative weight of the digital distractions over the non-digital ones.

Next, we are going to comment on significant correlations between pairs of distractions: one a non-digital distraction and the other a digital distraction. Table 6 shows the correlation coefficient matrix of these variables. The correlation coefficients with statistical significance are highlighted in bold, distinguishing between the *p*-values lower than 0.05 and lower than 0.01. The non-digital distraction "Students who make repetitive movements (touching their hair, cracking their knuckles, tapping their fingers, tapping their pencil, etc.)" is correlated significantly with the distractions "Students using social networks on any device (smartphone, tablet, laptop, etc.)", and "Students web browsing on any device (smartphone, tablet, laptop, etc.) for off-task purposes", with a Pearson coefficient of 0.878 for both cases. It should be remarked that the students had to select as relevant the distractions, not only caused by themselves, but also caused by other classmates, as the outcome in the student's attention is similar. In this regard, the non-digital distraction "Students who make repetitive movements (touching their hair, cracking their knuckles, tapping their fingers, tapping their pencil, etc.)" and, particularly, the digital distractions "Students using social networks on any device (smartphone, tablet, laptop, etc.)", and "Students web browsing on any device (smartphone, tablet, laptop, etc.) for off-task purposes" are likely to be caused by a student different from the student distracted by them. The students are strongly influenced by the overall classroom environment, and if there are many students making an unsuitable use of technological devices, the rest of the students will be tempted to make it as well or, in any case, they will be less focused on the tasks of the laboratory class.

The non-digital distraction "Professor difficult to understand" is correlated significantly with the digital distraction "Sounds produced by any device (smartphone, tablet, laptop, etc.) for different reasons (calls, notifications, etc.)" with a Pearson coefficient of 0.933. Professors should take measures to avoid the sounds produced by devices as they can make students stop paying attention, similar to the situation in which the students found it difficult to follow the explanations of their professor. Additionally, it is obvious that professors must do their best to ease the understanding of their explanations, and to monitor students' attention so that they can take correcting measures as soon as possible.

The non-digital distraction "Problems with equipment (computers, keyboards, mice, etc., that work badly)" is correlated significantly with the digital distractions "Students playing games on any device (smartphone, tablet, laptop, etc.)" and "Sounds produced by any device (smartphone, tablet, laptop, etc.) for different reasons (calls, notifications)" with Pearson coefficients of 0.755 and 0.842, respectively. Moreover, the non-digital distraction "Students making unnecessary comments" is correlated with the digital distraction "Students listening to music on any device (mp3 player, smartphone, tablet, laptop, etc.)" with a Pearson coefficient of 0.871. The relation between distractions is clear given that they are not caused by the distracted student but by other students, or situations, in the laboratory. This emphasizes the importance of a suitable environment for the student to have a satisfactory performance in the laboratory classes.

Two other significant correlations were found between non-digital and digital distractions. The non-digital distraction "Students who make repetitive movements (touching their hair, cracking their knuckles, tapping their fingers, tapping their pencil, etc.)" is correlated significantly with the digital distraction "Students using messaging applications for off-task purposes" with a Pearson coefficient of 0.874 and, finally, the non-digital distraction "Students wearing unusual clothes" is correlated significantly with the digital distraction "Students listening to music" with a Pearson coefficient of 0.848.

Sustainability **2023**, *15*, 604417 of 22**Table 6.** Correlation coefficient matrix of the non-digital and the digital distractions. Each row has a non-digital distraction and each column has a digital distraction.| | Students Playing Games on Any Device | Students Listening to Music | Students Using Social Networks | Students Using Messaging Applications for Off-Task Purposes | Students Using Email for Off-Task Purposes | Students Web Browsing for Off-Task Purposes | Sounds Produced by Any Device |
|---|---|---|---|---|---|---|---|
| Students talking to others | 0.709 | 0.093 | 0.511 | 0.410 | 0.588 | 0.585 | 0.396 |
| Students asking non-important questions | −0.641 | 0.046 | 0.197 | 0.123 | 0.000 | 0.010 | −0.722 |
| Students making unnecessary comments | 0.191 | 0.871 * | 0.523 | 0.421 | 0.580 | 0.431 | 0.268 |
| Students working on homework for other courses | 0.015 | −0.043 | 0.431 | 0.306 | 0.315 | 0.358 | −0.240 |
| Students wearing unusual clothes | 0.105 | 0.848 * | 0.474 | 0.384 | 0.652 | 0.074 | −0.100 |
| Students who make repetitive movements | 0.217 | 0.243 | 0.878 ** | 0.874 * | 0.719 | 0.878 ** | −0.238 |
| Symptoms of illness of a student | −0.498 | −0.138 | 0.204 | 0.202 | −0.010 | 0.137 | −0.402 |
| Students leaving early or arriving late | 0.457 | 0.167 | 0.312 | 0.184 | 0.312 | 0.469 | 0.099 |
| Students leaving or entering | −0.106 | 0.311 | 0.403 | 0.460 | 0.263 | 0.468 | 0.174 |
| Students eating or drinking | −0.417 | −0.162 | 0.042 | 0.205 | −0.139 | 0.079 | −0.083 |
| Students yawning, nodding, or falling asleep | −0.150 | 0.467 | 0.330 | 0.099 | 0.339 | 0.089 | −0.285 |
| Professor difficult to understand | 0.556 | −0.008 | −0.349 | −0.379 | −0.143 | −0.131 | 0.933 ** |
| Professor doing unusual or repetitive movements | −0.214 | 0.412 | 0.153 | −0.016 | 0.277 | −0.286 | −0.495 |
| Professor utilizing repetitive phrases or words | 0.308 | 0.357 | 0.659 | 0.567 | 0.734 | 0.402 | −0.119 |
| Furniture | 0.327 | 0.152 | −0.018 | −0.204 | 0.021 | 0.188 | 0.467 |
| Problems with equipment | 0.755 * | 0.061 | 0.082 | −0.030 | 0.250 | 0.272 | 0.842 * |
| Climate (too hot/cold) | 0.057 | −0.666 | −0.490 | −0.380 | −0.523 | −0.197 | 0.482 |

\* Correlation is significant at *p* < 0.05 level; ** Correlation is significant at *p* < 0.01 level.

On the other hand, Table 7 shows the correlation coefficient matrix of the digital distractions. The correlation coefficients with statistical significance are highlighted in bold, distinguishing between the *p*-values lower than 0.05 and lower than 0.01. There are also many significant correlations between pairs of digital distractions, for instance, among the distraction "Students using social networks on any device (smartphone, tablet, laptop, etc.)" with the distractions "Students using messaging applications on any device (smartphone, tablet, laptop, etc.) for off-task purposes", "Students using email (read or send) on any device (smartphone, tablet, laptop, etc.) for off-task purposes", and "Students web browsing on any device (smartphone, tablet, laptop, etc.) for off-task purposes". It goes without saying that a student who is prone to get distracted by some digital distraction is also prone to get distracted with other digital distractions; thus, his/her performance in the laboratory will decrease appreciably. The professor must develop strategies to minimize the influence of digital distractions on the performance of this kind of student.

Table 7. Correlation coefficient matrix of the digital distractions.

|  | Students Playing Games on Any Device | Students Listening to Music | Students Using Social Networks | Students Using Messaging Applications for Off-Task Purposes | Students Using Email for Off-Task Purposes | Students Web Browsing for Off-Task Purposes | Sounds Produced by Any Device |
|---|---|---|---|---|---|---|---|
| Students playing games on any device | 1 | 0.087 | 0.320 | 0.283 | 0.471 | 0.568 | 0.744 |
| Students listening to music |  | 1 | 0.619 | 0.474 | 0.727 | 0.302 | −0.014 |
| Students using social networks |  |  | 1 | 0.964 ** | 0.951 ** | 0.860 * | −0.131 |
| Students using messaging applications for off-task purposes |  |  |  | 1 | 0.875 ** | 0.873 * | −0.612 |
| Students using email for off-task purposes |  |  |  |  | 1 | 0.773 * | 0.027 |
| Students web browsing for off-task purposes |  |  |  |  |  | 1 | 0.191 |
| Sounds produced by any device |  |  |  |  |  |  | 1 |

\* Correlation is significant at *p* < 0.05 level; ** Correlation is significant at *p* < 0.01 level.

## 6. Conclusions

Social media is fully integrated into college students' daily lives. Today's university students spend significant amounts of time using their portable devices including smartphones, tablets, and laptops.

The present study contributes to the research into the impact of the use of technology on students in higher education by identifying the main digital distractors, especially during the development of laboratory classes in university campuses. The identification of digital distractors has been carried out from the point of view of students.

The students felt that the influence of digital distractions was comparable to that of non-digital distractions, even though the students were presented with three times as many non-digital distractions as digital distractions. The year of each subject does not particularly affect the influence of digital distractions with respect to the non-digital distractions. Moreover, more than two thirds of the students in the different subjects included in the study thought that their performance in the laboratory classes could be improved. This result was relevant in all subjects, and no differences in student perception were observed in the advanced course subjects. Distractions have a strong influence in poor self-perceived performance, and professors must do their best to improve that situation as soon as possible, preferably in subjects from the first years. In doing so, students can improve their performance in that subject and from then onwards in the subjects of the following years, easing the learning process in the crucial in-person classes, such as the laboratory classes.

The non-digital distractions that were the most highlighted by students were the distractions caused by other students, by the professor, or related to the laboratory equipment and climate conditions. All these distractions have in common that the distracted students did not cause the distraction. Other agents, such as other students or the professor, or an environmental factor, caused the distractions. Differently, the most prominent digital distractions, such as the use of social networks or messaging applications, are commonly caused by the distracted students, and neither another student nor an environmental factor. In this regard, awareness measures and self-control against these distractions can have a more direct impact on the reduction of their influence, thus improving students' performance in the laboratory classes.

The statistical study computing bivariate correlations produced illustrative results. The digital distraction "Students web browsing on any device (smartphone, tablet, laptop, etc.) for off-task purposes" was the only one correlated significantly with the question "How many students think they could better use their time during laboratory classes?", thus both stating the highlighted influence of this digital distraction on performance during the laboratory classes, and emphasizing the importance of adopting steps to avoid the ways in which students can be affected by this distraction and, as a result, work better in the laboratory classes. On the other hand, this digital distraction is correlated with the following digital distractions: "Students using social networks on any device (smartphone, tablet, laptop, etc.)", "Students using messaging applications on any device (smartphone, tablet, laptop, etc.) for off-task purposes", and "Students using email (read or send) on any device (smartphone, tablet, laptop, etc.) for off-task purposes". There are several digital distractions that can influence students, and usually students are driven to not only one of these distractions, but they are likely to be influenced by others.

Monitoring technology use and managing distractions are crucial to sustainable and effective use of technology in education. In fact, it is important to make students aware of the distractions caused by technology because media multitasking does not make a student more efficient; instead it is a bad habit. Student self-regulation around technology needs to be encouraged and professors should take this into account when designing instruction and planning academic activities. The presence of technology in college classrooms is not a problem itself. Rather, it is the way in which technology is incorporated into a scenario in which students are already prone to pay attention to too many things. Students are susceptible to distractions because they are not fully engaged in learning. Motivation plays a very important role, since a motivated student will be more focused on his or her task, avoiding distractions as much as possible. Future efforts are needed in this area.

In addition, it is also important to establish appropriate channels so that all incidents can be properly and promptly communicated, especially those related to laboratory equipment and climate conditions, which can be easily corrected.

Finally, although a great majority of the literature focuses on the positive side of technology, it is also essential to expand the research on the understudied dark side of technology, including the distractions caused by technology in students. It is necessary that students control the digital overload rather than letting the digital overload control them.

**Author Contributions:** Conceptualization, methodology, validation, and supervision, M.Á.P.-J. and J.M.A.-P.; formal analysis, investigation, visualization, and writing—original draft preparation, M.Á.P.-J. and D.G.-O.; writing—review and editing, M.Á.P.-J., D.G.-O. and J.M.A.-P. All authors have read and agreed to the published version of the manuscript.

**Funding:** This research received no external funding.

**Institutional Review Board Statement:** Not applicable.

**Informed Consent Statement:** Informed consent was obtained from all subjects involved in the study.

**Data Availability Statement:** The data presented in this study are contained within the article.

**Acknowledgments:** This research has been developed within the framework of a teaching innovation project supported by the Vice-Rectorate for Teaching Innovation and Digital Transformation of the University of Valladolid, which is a large public University in central Spain. The involvement of professors and students in teaching innovation projects is a source of pride for the University of Valladolid.

**Conflicts of Interest:** The authors declare no conflict of interest.